\documentclass{iau}

\usepackage{aas_macros}
\usepackage{amssymb}
\usepackage{amsmath}
\usepackage{tikz,xcolor,graphicx}
\graphicspath{{images/}}
\usepackage{multirow}

\definecolor{lime}{HTML}{A6CE39}
\DeclareRobustCommand{\orcidicon}{%
        \begin{tikzpicture}
        \draw[lime, fill=lime] (0,0)
        circle [radius=0.16]
        node[white] {{\fontfamily{qag}\selectfont \tiny ID}};
        \draw[white, fill=white] (-0.0625,0.095)
        circle [radius=0.007];
        \end{tikzpicture}
        \hspace{-2mm}
}

\newcommand{\orcidVP}{\href{https://orcid.org/0000-0002-3031-062X}{\orcidicon}}
\newcommand{\orcidMBD}{\href{https://orcid.org/0000-0001-6080-1190}{\orcidicon}}
\newcommand{\orcidHB}{\href{https://orcid.org/0000-0002-1959-6946}{\orcidicon}}
\newcommand{\orcidAB}{\href{https://orcid.org/0000-0002-4674-0704}{\orcidicon}}
\newcommand{\orcidICZ}{\href{https://orcid.org/0000-0001-9478-5731}{\orcidicon}}
\newcommand{\orcidMH}{\href{https://orcid.org/0000-0002-2363-5522}{\orcidicon}}
\newcommand{\orcidEL}{\href{https://orcid.org/0000-0001-7144-4766}{\orcidicon}}
\newcommand{\orcidAW}{\href{https://orcid.org/0000-0002-7501-9801}{\orcidicon}}

\newcommand{\trh}[1][]{\tau_\mathrm{rh#1}}

\newcommand{\Rh}[1][]{R_\mathrm{h#1}}
\newcommand{\Msun}{M_\odot}

\newcommand{\Rlim}{R_\mathrm{lim}}

\newcommand{\Nbin}{N_\mathrm{bin}}

\newcommand{\der}{\mathrm{d}}
\newcommand{\CRD}{C}

\begin{document}

\jnlPage{1}{4}
\jnlDoiYr{2025}
\volno{398 / MODEST-25}
\doival{} 

\aopheadtitle{\textbf{To appear in an edited form in:} Proceedings IAU Symposium}
\editors{Hyung Mok Lee, Rainer Spurzem and Jongsuk Hong, eds.}

\title{Insights into spatial mixing of multiple populations in~dynamically-young globular clusters}

\lefttitle{Pavl\'ik {\it et al.}}
\righttitle{Insights into spatial mixing of multiple populations in dynamically-young GCs}

\author{%
V\'aclav Pavl\'ik$^{1,2,\bigstar}$\orcidVP,
Melvyn B.~Davies$^{3}$\orcidMBD,
Ellen I.~Leitinger$^{4,5}$\orcidEL,
Holger Baumgardt$^{6}$\orcidHB,
Alexey Bobrick$^{7,8,9}$\orcidAB,
Ivan Cabrera-Ziri$^{10}$\orcidICZ,
Michael Hilker$^{5}$\orcidMH, and
Andrew J.~Winter$^{11,12,13}$\orcidAW
}

\affiliation{%
$^1$ Astronomical Institute, Czech Academy of Sciences, Bo\v{c}n\'i~II~1401, 141~00~Prague~4, Czech Republic\\
$^\bigstar$ \email{pavlik@asu.cas.cz}\\
$^2$ Dept.~of Astronomy, Indiana University, Swain Hall West, 727 E 3$^\text{rd}$ St., Bloomington, IN 47405, USA\\
$^3$ Centre for Mathematical Sciences, Lund University, Box 118, SE-221 00 Lund, Sweden\\
$^4$ Dip.~di Fisica e Astronomia, Univ.~degli Studi di Bologna, Via Gobetti 93/2, I-40129 Bologna, Italy\\
$^5$ European Southern Observatory, Karl-Schwarzschild-Str. 2, D-85748 Garching, Germany\\
$^6$ School of Mathematics and Physics, The University of Queensland, St. Lucia, QLD 4072, Australia\\
$^7$ School of Physics and Astronomy, Monash University, Clayton, Victoria 3800, Australia\\
$^8$ ARC Centre of Excellence for Gravitational Wave Discovery -- OzGrav, Australia\\
$^9$ Technion -- Israel Institute of Technology, Physics Department, Haifa, Israel 32000\\
$^{10}$ Vyoma GmbH, Karl-Theodor-Straße 55, 80803 Munich, Germany\\
$^{11}$ Max-Planck Institute for Astronomy (MPIA), K\"onigstuhl 17, 69117 Heidelberg, Germany\\
$^{12}$ Univ.~C\^ote d'Azur, Observatoire de la C\^ote d'Azur, CNRS, Laboratoire Lagrange, 06300 Nice, France\\
$^{13}$ Astronomy Unit, School of Physics and Astronomy, Queen Mary Univ.\,of London, London E1\,4NS, UK
}

\begin{abstract}
Many galactic globular clusters (GCs) contain at least two stellar populations. Recent observational studies found that the radial distributions of the first (P1) and second population (P2) differ in dynamically-young GCs. Since P2 is conventionally assumed to form more centrally concentrated, the rapid mixing (or even inversion) in some GCs but not others is puzzling. We investigate whether dynamical processes specific to certain GCs might cause this. Specifically, we evaluate the expansion of P2 by binary-single interactions in the cluster core and whether these can mix the P1/P2 radial distributions, using a set of toy-models with varying numbers and masses of primordial binaries. We find that even one massive binary star can push the central P2 outwards, but multiple binaries are required to fully mix P1 and P2 within a~few relaxation times. We also compare our results to observed properties of mixed young GCs (NGC 4590, 5053, or 5904).
\end{abstract}


\maketitle

\section{Introduction}

Massive and old globular clusters (GCs) are known to host multiple stellar populations with distinct chemical compositions and kinematics \citep[e.g.][]{carretta_etal2010, gratton_review2012, multipop_review18,vesperini_etal21,leitinger_etal24}. The first population, or P1 stars, shows primordial abundances, while the enriched second population, P2, exhibits light-element variations that imply complex formation scenarios—ranging from \textit{in-situ} formation in enriched gas to accretion of processed material \citep[e.g.][]{carretta_etal2010, bekki_2011}. Many formation models suggest P2 stars form preferentially in the core \citep[e.g.][]{lacchin_etal22}, but observations of dynamically-young GCs show diverse spatial distributions: some retain a centrally concentrated P2 population, others have higher numbers of P1 in the core, some are fully radially mixed \citep{leitinger_etal23}, see also a selection of Galactic GCs in Fig.~\ref{fig:Ap_obs}.

This variety in radial distributions of P1 and P2 suggests that some internal dynamics shaped the present-day structure of GCs.
Since binary--single star scattering is a prevalent cause of stellar population heating and ejections \citep[see, e.g.][]{heggie_hut, binney_tremaine}, we test whether massive binaries in GC cores could drive early redistribution or inversion of P2 stars.

\section{Methods}

To gain insights into the mixing capabilities of massive binary stars inside star clusters, we study a set of toy-models --- i.e.~our initial conditions are intentionally simplified to isolate the effects of the binary--single star interactions in the cluster core. In our follow-up paper \citep{pavlik_etal_scbh}, we expand this study to more realistic star clusters of various masses, radial profiles, and primordial binary star properties. Specifically, here we initialise all stars on circular orbits, so that their radial motion early in the evolution would be mostly affected by the binary--single star interactions and not by the random orientation of their orbits. We also adopt a uniform density distribution of stars to filter out rapid relaxation processes in the core.

Each cluster has ten thousand equal mass stars with $1\,\Msun$ and either zero, one, or ten binaries with $30\,\Msun$ components, circular orbits, and semi-major axes of 5\,au.
The masses of binaries are derived from the typical gravitational-wave sources, assuming they would be represented by black holes in real GCs. We label the single low-mass stars by their initial positions as `P2' (below the quarter-mass radius) and `P1' (everything else). We do not count the binaries towards either population. We average our results over ten random realisations of each model.

\section{Results}

\begin{figure*}
    \centering
    \includegraphics[width=0.255\linewidth]{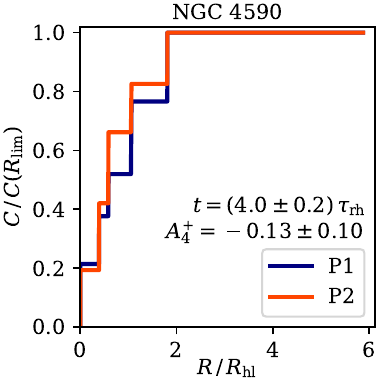}
    \includegraphics[width=0.242\linewidth]{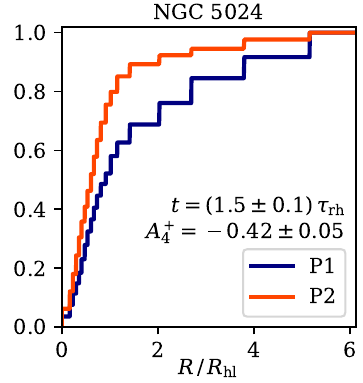}
    \includegraphics[width=0.242\linewidth]{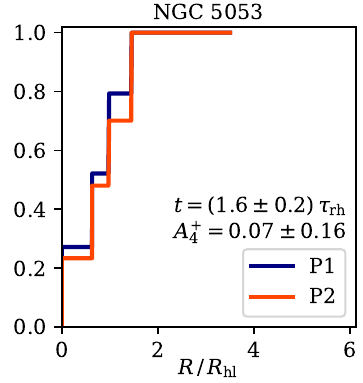}\\
    \includegraphics[width=0.242\linewidth]{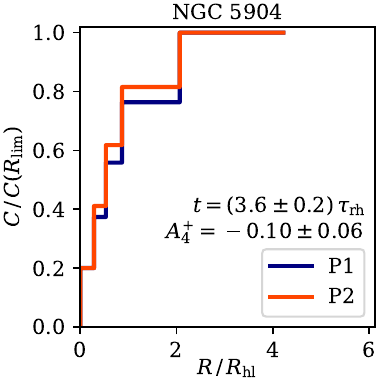}
    \includegraphics[width=0.242\linewidth]{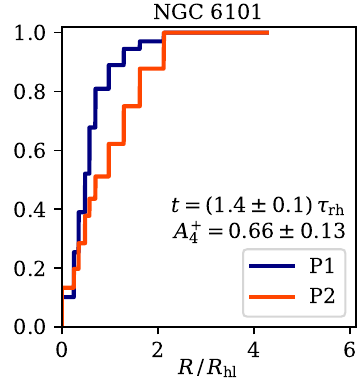}
    \includegraphics[width=0.242\linewidth]{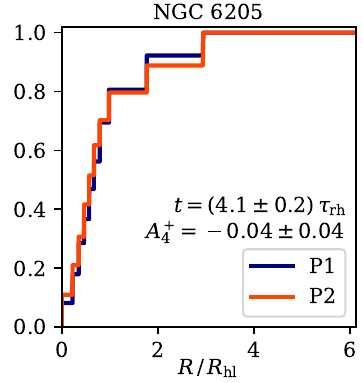}\\
    \caption{Cumulative distribution functions of P1 and P2 stars in a sample of dynamically-young Galactic GCs. We note that the displayed radial range goes up to six half-light radii, $\Rh[l]$, to show the whole clusters, but the parameter $A^+_4$ is calculated from stars only up to $4.27\,\Rh[l]$ \citep[data taken from][]{leitinger_etal24}.}
    \label{fig:Ap_obs}
\vspace{2\floatsep}
    \centering
    \includegraphics[width=0.255\linewidth]{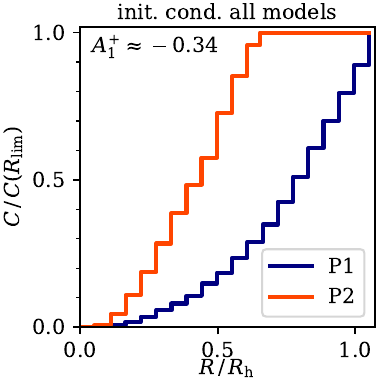}
    \includegraphics[width=0.242\linewidth]{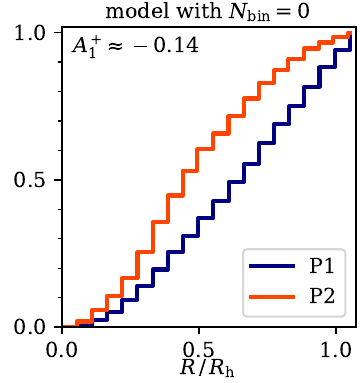}
    \includegraphics[width=0.242\linewidth]{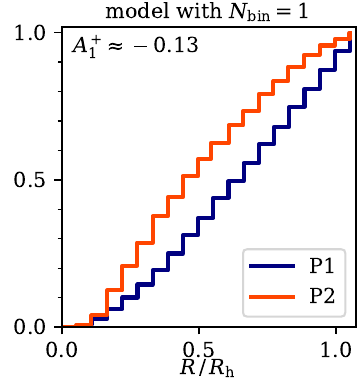}
    \includegraphics[width=0.242\linewidth]{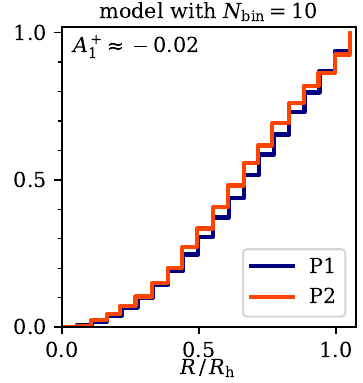}\\
    \includegraphics[width=0.255\linewidth]{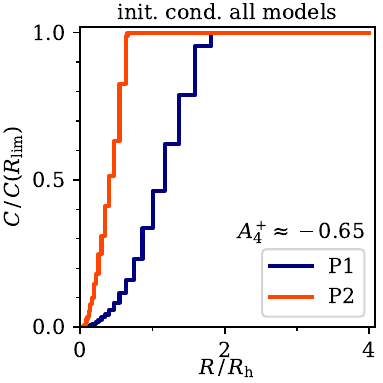}
    \includegraphics[width=0.242\linewidth]{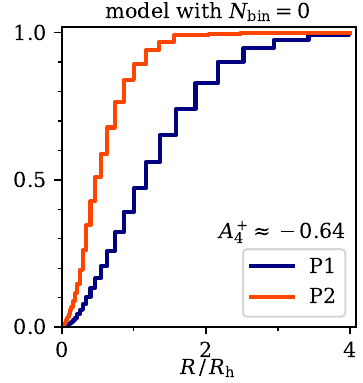}
    \includegraphics[width=0.242\linewidth]{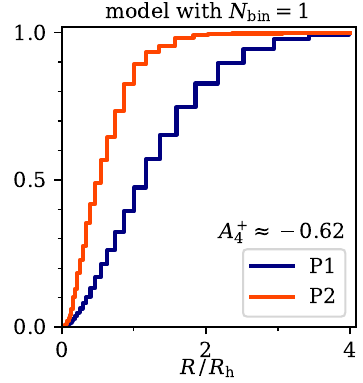}
    \includegraphics[width=0.242\linewidth]{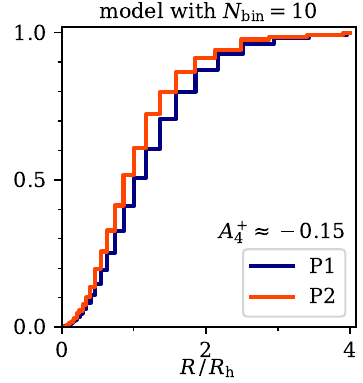}\\
    \includegraphics[width=0.255\linewidth]{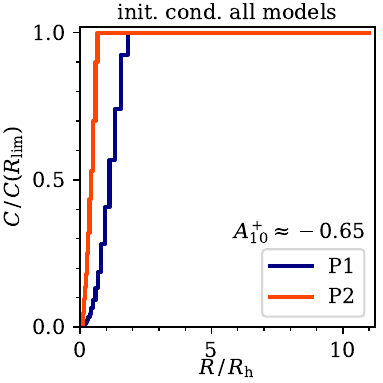}
    \includegraphics[width=0.242\linewidth]{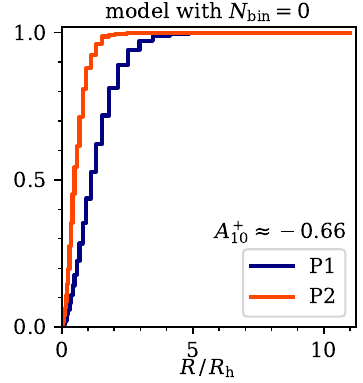}
    \includegraphics[width=0.242\linewidth]{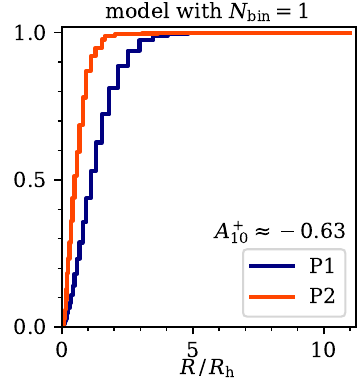}
    \includegraphics[width=0.242\linewidth]{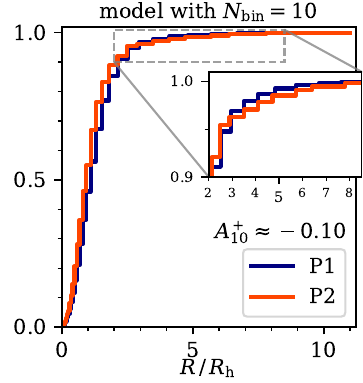}\\
    \caption{Cumulative distribution functions of P1 and P2 stars and their corresponding $A^+_{R{\rm lim}}$ values in our cluster models. Each row corresponds to a different projected radial range, $\Rlim$. The left-hand column shows the initial conditions, the columns on the right each show the evolved models (each containing a different number of binaries) at $2\,\trh$.}
    \label{fig:Ap_models}
\end{figure*}

We study the redistribution of P1 and P2 using their projected cumulative radial distribution functions, $\CRD(R)$. From these we calculate the $A^+$ parameter \citep{alessandrini_etal16}
\begin{equation}
    \label{eq:Ap}
    A^{+}_{R\mathrm{lim}}
    = \frac{1}{\Rh}
      \int_0^{\Rlim}{\left[ \frac{\CRD_\mathrm{P1}(R)}{\CRD_{\rm P1}(\Rlim)} -
                            \frac{\CRD_\mathrm{P2}(R)}{\CRD_{\rm P2}(\Rlim)} \right] \der R} \,,
\end{equation}
where $\Rlim$ is the chosen limiting radius for normalisation. In Fig.~\ref{fig:Ap_models}, we use $\Rlim = 1\,\Rh$ for the central region, $4\,\Rh$ to compare with observations, and $10\,\Rh$ to include most of the cluster. In the models (Fig.~\ref{fig:Ap_models}), we scale $\Rlim$ by the projected half-mass radius, while the radial ranges of the observations (Fig.~\ref{fig:Ap_obs}) are scaled by the half-light radii, $\Rh[l]$; however, $\Rh = \Rh[l]$ in equal-mass star clusters.
The dynamical age of the modelled and observed clusters is expressed in terms of their present-day half-mass relaxation time. In the modelled clusters with $N$ stars and the total mass $M$, this is in projection \citep[e.g.][]{spitzer_hart_relax, binney_tremaine}
\begin{equation}
    \label{eq:trh}
    \trh \approx \frac{0.138\,N}{\ln{(0.4 N)}} \sqrt{\frac{\Rh^3}{GM}} \,.
\end{equation}

In Fig.~\ref{fig:Ap_models}, we plot the initial conditions of our models in the left-hand column and their evolved states at $2\,\trh$ in the right-hand panels (to be comparable with the observed dynamically-young GCs). In the top row of this figure, we see that the initially separated populations fully mix in the region up to $1\,\Rh$ in the model with ten binaries ($A^+_1\approx-0.02$). The radial redistribution of P1 and P2 is lower in the models with one or no binary (and the difference between these two models is negligible). Even in the regions up to 4 or $10\,\Rh$ (middle and bottom row of the figure), the model with ten binaries appears mixed. We can even see a hint of inversion between $3$ and $8\,\Rh$ (see the zoomed-out plot in the bottom-right panel). We do not see mixing or any signs of inversion at these radial scales in the models with one or zero binaries.

We list all values of $A^+$ at the evolutionary times $2\,\trh$ and $4\,\trh$ in Tables~\ref{tab:Ap_2trh} and~\ref{tab:Ap_4trh}, respectively. From these and the figures above, we see that our simple models do not exhibit strong inversion ($A^+>0$) as in the case of the observed cluster NGC~6101. However, already at $2\,\trh$, the models with ten primordial binaries are comparable to the mixed GCs NGC~4590, 5053, 5904, or 6205. In turn, the models with a lower binary count are similar to the P2-centrally-concentrated NGC~5024. The level of P1/P2 mixing increases with the evolutionary time in our models \citep[cf.][]{leitinger_etal23,leitinger_etal24}. We may also compare the two tables and see that $A^+$ usually decreases, except for the model with ten binaries, where its value stays almost the same (within the uncertainties).

\begin{table}
    \begin{minipage}{0.495\linewidth}
        \centering
        \caption{The mean values of $A^+$ at $2\,\trh$\\ and their uncertainties (see also  Fig.~\ref{fig:Ap_models}).}
        \begin{tabular}{|c|ccc|}
            \hline
            $\Nbin$    & $A^+_1$ & $A^+_4$ & $A^+_{10}$ \\
            \hline
            0  & $-0.14$ & $-0.64$ & $-0.66$ \\
            1  & $-0.13$ & $-0.62$ & $-0.63$ \\
            10 & $-0.02$ & $-0.15$ & $-0.10$ \\
            \hline
            $\sigma$  & $\pm0.01$ & $\pm0.03$ & $\pm0.05$ \\
            \hline
        \end{tabular}
        \label{tab:Ap_2trh}
    \end{minipage}
    \hfill
    \begin{minipage}{0.495\linewidth}
        \centering
        \caption{The mean values of $A^+$ at $4\,\trh$\\ and their uncertainties.}
        \begin{tabular}{|c|ccc|}
            \hline
            $\Nbin$    & $A^+_1$ & $A^+_4$ & $A^+_{10}$ \\
            \hline
            0  & $-0.09$ & $-0.57$ & $-0.62$ \\
            1  & $-0.08$ & $-0.51$ & $-0.57$ \\
            10 & $-0.01$ & $-0.10$ & $-0.11$ \\
            \hline
            $\sigma$  & $\pm0.01$ & $\pm0.03$ & $\pm0.05$ \\
            \hline
        \end{tabular}
        \label{tab:Ap_4trh}
    \end{minipage}
\end{table}

\section{Conclusions}

Our preliminary numerical analysis shows that massive binary stars (presumably black holes) can cause mixing of stellar populations inside a star cluster. We investigate this mechanism further --- both theoretically and numerically --- in \citet{pavlik_etal_scbh}.

\begin{acknowledgements}
This research used computational resources from e-INFRA CZ (project ID:90254), supported by the Ministry of Education, Youth and Sports of the Czech Republic; and the computational cluster Virgo at the Astronomical Institute of the Czech Academy of Sciences.
V.P.~is funded by the European Union's Horizon Europe and the Central Bohemian Region under the Marie Skłodowska-Curie Actions -- COFUND, Grant agreement \href{https://doi.org/10.3030/101081195}{ID:101081195} (``MERIT''). V.P.~also acknowledges support from the project RVO:67985815 at the Czech Academy of Sciences.
E.I.L.~acknowledges support from the ERC Consolidator Grant funding scheme (project ASTEROCHRONOMETRY, \url{https://www.asterochronometry.eu}, G.A.~n.~772293).
A.B.~acknowledges support from the Australian Research Council (ARC) Centre of Excellence for Gravitational Wave Discovery (OzGrav), through project number CE230100016.
M.H.~acknowledges financial support from the Excellence Cluster ORIGINS which is funded by the Deutsche Forschungsgemeinschaft (DFG, German Research Foundation) under Germany’s Excellence Strategy – EXC 2094 – 390783311.
A.J.W.~has received funding from the European Union’s Horizon 2020 research and innovation programme under the Marie Skłodowska-Curie grant agreement No 101104656 and the Royal Society's University Research Fellowship, reference URF/R1/241791.
\end{acknowledgements}

\bibliographystyle{aa}
\bibliography{pavlik_modest25}

\end{document}